\documentclass[12pt]{article}
\usepackage[latin1]{inputenc}
\usepackage{amsfonts}
\usepackage{cite}
\usepackage{amsxtra}
\usepackage[margin=3cm]{geometry}
\usepackage{color}
\usepackage{hyperref}
\hypersetup{linktocpage=true}
\usepackage{graphicx}
\usepackage{mathrsfs}
\usepackage{multirow}
\usepackage{tikz}
\usepackage{amsmath,latexsym,amssymb,slashed}
\usepackage{bbm}
\usepackage{pdfpages}
\usepackage{centernot}
\usepackage{multirow}
\usepackage{varwidth}
\usepackage{slashbox}
\usepackage{amsthm}
\usepackage{xypic}

\usepackage[toc,page]{appendix}
\usepackage{bookmark}

\numberwithin{equation}{section}
\newcommand{\beq}{\begin{equation}}
\newcommand{\eeq}{\end{equation}}
\def\be {\begin{equation}}
\def\ee {\end{equation}}
\def\ba#1\ea{\begin{align}#1\end{align}}
\def\baed#1\eaed{\begin{aligned}#1\end{aligned}}
\def\bged#1\eged{\begin{gathered}#1\end{gathered}}
\def\bea{\begin{eqnarray}}
\def\eea{\end{eqnarray}}

\def\a{\alpha}
\def\b{\beta}

\def\ve{\varepsilon}

\def\F{\Phi}
\def\g{\gamma}
\def\G{\Gamma}

\def\L{\Lambda}
\def\m{\mu}
\def\n{\nu}
\def\o{\omega}
\def\O{\Omega}

\renewcommand{\t}{\theta}
\def\r{\rho}
\def\s{\sigma}

\def\z{\zeta}

\let\foo\bar
\renewcommand{\bar}[1]{ {\foo{  #1} }{} }

\newlength{\dhatheight}

\newcommand{\eq}[1]{\begin{equation}\begin{split}#1\end{split}\end{equation}}

\newcommand{\all}[1]{\begin{align*}#1\end{align*}}

\newcommand{\arxdg}[1]{\href{http://arxiv.org/abs/math/#1}{\tt math.dg/#1}}
\newcommand{\arxth}[1]{\href{http://arxiv.org/abs/hep-th/#1}{[{\tt hep-th/#1}]}}
\newcommand{\arx}[1]{[\href{http://arxiv.org/abs/#1}{\tt #1}]}

\newcommand{\acal}{\mathcal{A}}

\newcommand{\fcal}{\mathcal{F}}

\newcommand{\mcal}{\mathcal{M}}
\newcommand{\ncal}{\mathcal{N}}

\newcommand{\pcal}{\mathcal{P}}
\newcommand{\rcal}{\mathcal{R}}
\newcommand{\scal}{\mathcal{S}}

\newcommand{\rbb}{\mathbbm{R}}

\newcommand{\vt}{\vartheta}

\newcommand{\ea}{\bigwedge\nolimits^{\!\bullet} T^*}

\newcommand{\p}{\partial}

\def\d{\text{d}}

\newcommand{\ti}[1]{\tilde{#1}}

\newcommand{\cg}{\check{\gamma}}

\renewcommand{\ve}{\varepsilon}
\newcommand{\wt}[1]{\widetilde{#1}}

\numberwithin{equation}{section}

\begin{document}
%%%%%%%%%%%%%%%%%%%%%%%%%%%%%%%%%%%%%%%%%%%%%%%%

\baselineskip=16pt
\setlength{\parskip}{6pt}

\begin{titlepage}
\begin{minipage}{0.5\textwidth}
{\color{white} a}
\end{minipage}
\begin{minipage}{0.5\textwidth}
\vspace*{-.65\baselineskip}
\flushright IPhT-17/165
\end{minipage}

\begin{center}

\vspace*{1.7cm}

{\LARGE \bf  Ten-dimensional lifts of global supersymmetry on curved spaces
}

\vskip 1.6cm

\renewcommand{\thefootnote}{}

\begin{center}
 \normalsize
 Ruben Minasian$^{\,a,b}  $ and Dani\"el Prins$^{\,a,c,d}$
\end{center}
\vskip 0.1cm
 {\sl $^a$ Institut de physique th\'eorique, Universit\'e Paris Saclay, CNRS, CEA \\F-91191 Gif-sur-Yvette,
France}
\vskip 0.1cm
{\sl $^b$ School of Physics, Korea Institute for Advanced Study, Seoul 130-722, Korea}
\vskip 0.1cm
{\sl $^c$ Dipartimento di Fisica, Universit\`{a} di Milano-Bicocca,\\
 I-20126 Milano, Italy}
 \vskip 0.1cm
{\sl $^d$ INFN, sezione di Milano-Bicocca, I-20126 Milano, Italy}
\vskip 0.2 cm
{\textsf{ruben.minasian@cea.fr}} \\
{\textsf{daniel.prins@cea.fr}}\\
\end{center}

\vskip 1.5cm
\renewcommand{\thefootnote}{\arabic{footnote}}

\begin{center} {\bf ABSTRACT } \end{center}
Admissible curved space backgrounds for four-dimensional supersymmetric field theories are determined by solving Killing spinor equations of four-dimensional off-shell supergravities.
These can be obtained by combining ten-dimensional type IIB supersymmetry with D-brane kappa-symmetry and identifying auxiliary fields of the four-dimensional supergravity fields in terms of type IIB fields.
In this paper we show how to extend a number of solutions of four-dimensional Killing spinor equations with four or less supercharges to solutions of the ten-dimensional supersymmetry constraints.

\end{titlepage}

\newpage		
\tableofcontents
\vspace{20pt}

\setcounter{page}{1}
\setlength{\parskip}{9pt}

\newpage

\section{Introduction}
Much recent work  was devoted to constructing supersymmetric field theories  in four dimensions on non-trivial Riemannian manifolds, with four or less supercharges \cite{fs, st, ktz, dfs,shamir, df, cmktz, liu, js}. The field theory is generally coupled to an $\ncal=1$ off-shell supergravity which is then  treated as a background.  Important constraints arise from demanding that the fermion, i.e. gravitino, variations vanish for the supergravity. The Killing spinor equations, obtained this way, are then solved by a specific Killing spinor, metric, and a profile for the appropriate auxiliary fields which are determined by the choice of supergravity (generally, either old minimal or new minimal). The number of Killing spinors with specific chirality which solve the Killing spinor equations for given metric and auxiliary fields determines the amount of supersymmetry under which the field theory is invariant. This procedure has  lead to a list of admissible four-manifolds that support globally supersymmetric gauge theories.

In \cite{triendl} it was shown how backgrounds which  simultaneously obey the supersymmetry conditions for bulk type IIB theory and D3-branes
imply the vanishing of the gravitino variation of a four-dimensional $\ncal =1$ off-shell supergravity.\footnote{See \cite{mrs} for a  discussion focusing on M5-branes.} However, the four-dimensional supergravity appearing here is the so-called  16/16 formulation \cite{16161, 16162, 16163}, and can be thought of  either as old minimal supergravity coupled to a vector multiplet or new minimal supergravity coupled to a chiral multiplet. The four-dimensional $\ncal =1$ Killing spinor equations  are thus formally reproduced with  the auxiliary fields of the supergravity being replaced by combinations of the fluxes, the dilaton and components of the spin connection of ten-dimensional type IIB theory. On the other hand, the desired four-dimensional Killing spinor equations form only a small subset  of the IIB closed and open string supersymmetry conditions.\footnote{Non-trivial worldvolume flux is associated with non-linearly realised supersymmetries of the field theory \cite{mpt}. Here, we are interested in the simplest solutions and we will be setting it to zero.}

In this paper, we will consider the full set of string supersymmetry conditions, and demonstrate how to obtain four-dimensional curved Riemannian manifolds as part of ten-dimensional supersymmetric configurations. Given a four-dimensional background, consisting of a four-dimensional metric and auxiliary fields, we wish to find a ten-dimensional string background that satisfies the supersymmetry equations, such that the auxiliary fields are specified in terms of the IIB fields. In addition, we wish to identify the supercharges preserved in the given four-dimensional background as the four-dimensional component of the ten-dimensional Killing spinors.

By locally splitting ten-dimensional spacetime into directions parallel and orthogonal to the D3-brane,  in the set of all string supersymmetry conditions we can separate the four-dimensional  equations  from the rest. Then, introduction of a pair of D7-branes breaks the four-dimensional supersymmetry from $\ncal=4$ to $\ncal =1$.
We will see how  the string supersymmetry conditions imply the vanishing of the variations of the gravitino  (see \eqref{extgrav})  as well as the additional spin 1/2 fermion of 16/16 supergravity (see \eqref{dilatino}). In view of its stringy origin, we will refer to this second equation as the (modified) dilatino equation. There are also  a number of equations which have no four-dimensional interpretation (see \eqref{intgrav}), originating from the internal (six-dimensional) gravitino variation. These three sets of equations combined are equivalent to the IIB supersymmetry conditions, taking into account that a number of components of the fluxes have been set to zero. As explained in section \ref{sec:setup}, all terms appearing in these equations have explicit expressions in terms of ten-dimensional  fields.

A specific known solution of the four-dimensional $\ncal =1$ Killing spinor equations can then be examined with the purpose of extending it to a full solution of ten-dimensional supersymmetry conditions. This procedure will be elaborated in section \ref{sec:4d}.

The requirements imposed in ten dimensions follow the guidelines from the four-dimensional backgrounds.
In the four-dimensional analysis the equations of motion for the auxiliary fields are not imposed, and it is not required to solve supergravity theories. Hence one does not expect that the ten-dimensional lifts will generically involve fluxes that satisfy Bianchi identities. In this paper we are only interested in solutions of ten-dimensional supersymmetry conditions. In addition, while the general expectation is that the auxiliaries are real, this is no longer the case once one considers Euclidean four-dimensional supergravities.\footnote{Four-dimensional Euclidean supergravities are not well-studied. See \cite{reys} for a study of Euclidean supergravity arising from a timelike reduction from five dimensions.}
Hence it is not too surprising that the uplifting some of the four-dimensional backgrounds may lead to the appearance of the complex formulation of ten-dimensional supersymmetry \cite{bergshoeff}.  The complexified theory encompasses the  so-called variant supergravities in ten dimensions with different spacetime signatures, notably  Euclidean type IIA (but not IIB) theory, which have been predicted by dualities \cite{hull}.  An appropriate choice of the reality condition may yield all existing variants.  As will be explained in section \ref{sec:complex}, for four-dimensional theories which for a given chirality preserve only a single supercharge, the only possible ten-dimensional lifts are to the complexified theories.

Our solutions can be grouped into two classes, presented respectively in sections  \ref{sec:simple} and \ref{sec:complex}. The first class involves solutions with at least two preserved supercharges of the same chirality, and is based on a simple ansatz for solving the dilatino variation. All solutions in this class satisfy the supersymmetry conditions of the standard IIB supergravity. The four-dimensional manifolds appearing in this class are noncompact and the examples include hyperbolic spaces ($H_4$ and $H_3\times S^1$) and  $S^3 \times \rbb$ and are reviewed in section \ref{sec:simple}.  The second class, exemplified by $S^4$ and $S^3 \times S^1$, involves solutions to the complexified supersymmetry variations. These solutions are presented in section \ref{sec:complex}. In addition, the details of a general $S^4$ solution can be found in appendix \ref{s4calc}.\footnote{Some details on Killing spinor equations in minimal and 16/16 supergravities can be found in appendix \ref{cd}. Appendix \ref{connections} contains some useful formulae relating ten and four-dimensional spin connections.}

In spite of recent progress in the study of gauge theory on curved spaces, many important questions remain open \cite{pestun}. Our goal was to show that ten-dimensional lifts of four-dimensional Killing spinor equations exist, not to be exhaustive in our search for such lifts, and we have made several simplifying assumptions. The sampling of solutions presented here involves setting to zero as many fields as possible, such as dropping half of the equations as explained in the commentary to table \ref{tab1}.
We have verified that not only the four-dimensional Killing spinor equations can be found within the combination of ten-dimensional bulk and D-brane supersymmetry variations, but that actual solutions of the former can be lifted to the solutions of the later. A choice of ten-dimensional metric and fluxes  results in a choice of an admissible Riemann manifolds and global symmetry background for a four-dimensional supersymmetric field theory. For a given four-dimensional Riemannian manifold, there is no unique choice of auxiliary fields, hence one does not expect a unique uplift. Moreover, we show that even for a given choice of auxiliary fields the uplift need not be unique. However, the ten-dimensional perspective should hopefully prove to be useful in further study of these gauge theories.

\section{The set up}\label{sec:setup}
In order to determine which manifolds can support globally supersymmetric field theories,  a number of four-dimensional equations need to be solved. We will first consider the derivation of this set of four-dimensional equations in terms of ten-dimensional fields.

The type IIB supersymmetry conditions, i.e. the vanishing of the ten-dimensional gravitino and dilatino respectively,  in ten dimensions are given by
\eq{
\delta_{\hat \varepsilon} \Psi_M &=  \hat{D}_M \hat{\ve}
= \nabla_M \hat \varepsilon + \tfrac18 H_{MNP} \Gamma^{NP} {\cal P} \hat \varepsilon + \tfrac{1}{16} e^{\phi} \sum_n \tfrac{1}{(2n-1)!} F_{M_1 \dots M_{2n-1}} \Gamma^{M_1 \dots M_{2n-1}} \Gamma_M {\cal P}_{n} {\hat \varepsilon}  \\
\delta_{\hat \varepsilon} \chi &= \hat{D} \hat{\varepsilon} = ( (\partial_M \phi)\Gamma^M {\hat \varepsilon} +\tfrac{1}{12} H_{MNP} \Gamma^{MNP}{\cal P}) {\hat \varepsilon} - \tfrac18 e^{\phi} \sum_n  \tfrac{6-2n}{(2n-1)!}  F_{M_1 \dots M_{2n-1}} \Gamma^{M_1 \dots M_{2n-1}} {\cal P}_{n} {\hat \varepsilon}
}
Capital latin letters will be used to denote ten-dimensional indices, those from the beginning of the alphabet ($A,B, ...= 0,...9$) being reserved for the tangent space indices, while those from the middle ($M,N, ...=0,...,9$) for the curved ones.

In addition, the condition for a supersymmetric Euclidean D3-branes wrapping a manifold $\scal$ placed in a type IIB background balancing the worldvolume supersymmetry and kappa-symmetry is given by
\eq{\label{kappa}
\G_0 {\hat \ve} = i \g_{(4)} \pcal_1 \hat{\ve} = {\hat \ve} \;.
}
Here, we have defined
\begin{equation}
{\cal P}   = \left(\begin{array}{cc} 1 & 0 \\ 0 & -1 \end{array}\right) \ , \qquad
{\cal P}_n = \left(\begin{array}{cc} 0 & 1 \\  (-1)^{n} & 0 \end{array}\right)
\end{equation}
as generators of the $SL(2, \rbb)$ S-duality. The tangent space and curved indices on $\scal$ are denoted respectively by middle of the alphabet latin ($m, n, ... = 1,...,4$) and greek ($\m, \n, ... = 1, ...,4$) letters.\footnote{We work with four-component Weyl spinors. We will never write out spinor indices, except in appendix \ref{cd} where we demonstrate how our conventions are related to those in the literature, where everything will be four-dimensional; hopefully the use of $\a, \dot{\a}$ for spinor indices there will cause no confusion.} On $\mcal_6$, we denote tangent space  indices by letters from the beginning of the latin alphabet ($a, b, ... = 0,5,...,9$) and curved space indices by greek letters ($\a, \b, ... = 0,5, ...,9$).

Note that we have taken a special form of projection in \eqref{kappa} where  the worldvolume flux $\fcal_{mn} = e_{[m}^M e_{n]}^N B_{MN} + 2 \pi \a' F_{mn}$ has been set to zero. The resulting  restrictions on the allowed values for the background NSNS three-form $H$ will be taken into consideration.  The more general situation with non-trivial worldvolume flux has been discussed in \cite{mpt}. We  impose this constraint in order to not have to deal with non-linear supersymmetry for the background since our interest is in finding the ten-dimensional lifts of four-dimensional Riemannian manifolds $\scal$ capable of supporting globally supersymmetric gauge theories.

We think of $\scal \subset \mcal_{10}$, wrapped by the Euclidean D3-brane,  as a submanifold of the ten-dimensional spacetime.
All our analysis is local, and we work with a split ten-dimensional tangent bundle. As we shall see in our solutions the spacetime metric is a warped product $\mcal_{10} = \scal \times \mcal_6$. We can construct more complicated metrics that satisfy the supersymmetry conditions, but for our purposes this will not be necessary; we will restrict our attention to finding the simplest possible solutions.
We will refer to $\scal$ as the external space and to the Lorentzian space $\mcal_6$ as the internal space.

We will denote by $\ve =  (\ve^1,\ve^2)$ the pullback of the ten-dimensional Killing spinor doublet ${\hat \ve}$ to the brane, which we locally decompose  as
\eq{\label{ksdecomp}
\ve^1 = \xi^+_j \otimes \eta^+_j + \xi^-_j \otimes \eta^-_j + \text{c.c.} \;,
}
with $\ve^2$ determined by the D3-brane supersymmetry condition \eqref{kappa}. The pseudoreal chiral spinors $\xi^\pm_j$ of $Spin(1,5)$ will be referred to as the `internal spinors', and the pseudoreal chiral spinors $\eta^\pm_j$ of $Spin(4)$ as the `external spinors'. They can be considered as the spinors respectively perpendicular and parallel to the brane. We should stress that there is absolutely no guarantee that such a decomposition holds globally, nor that this is even possible for generic $\scal$, since in general $\scal$ need not admit a spin structure.
Relevant examples for which globally well-defined nowhere vanishing four-dimensional chiral Killing spinors fail to exist are generic K\"{a}hler manifolds and $S^4$. We refer to the literature \cite{dfs} for details on how to deal with such issues.

A priori, the index in \eqref{ksdecomp} satisfies $j\in \{1,2\}$ since we have 32 supercharges. We are, however, interested in four-dimensional backgrounds preserving four or less supercharges; the superfluous symmetry will be removed by introducing supersymmetric D7-branes in the next section. For the rest of this section, we will consider the more supersymmetric case.

The combined open and closed string supersymmetry conditions are equivalent to the vanishing of each of the following terms:
\eq{\label{n=4}
\begin{alignedat}{4}
L_m &= \frac12 \{D_m, \G_0 \} \ve && \qquad \qquad &&  R_m  &&= \frac12  [D_m, \G_0]  \ve \\
L_a &= \frac12 [D_a, \G_0]    \ve             && \qquad \qquad &&  R_a  &&= \frac12 \{D_a, \G_0\} \ve \\
L   &= \frac12  [D - \G^a D_a, \G_0] \ve && \qquad \qquad &&   R   &&= \frac12 \{D - \G^a D_a, \G_0\}    \ve \;.
\end{alignedat}
}
As can be seen from table \ref{tab1}, with the exception of the ten-dimensional metric which dictates all four spin connection components with mixed indices, the sets of fields appearing in the $L$ and $R$ terms are disjoint. Our interest lies in examining string backgrounds which in some sense embed four-dimensional supergravity backgrounds. Such four-dimensional backgrounds are determined by four-dimensional Killing spinor equations (we will be more specific later on), hence it is the $L$-terms, which are related to $\nabla_m^{(4)}$, that are of interest to us. Therefore, the simplest ansatz (which we will use throughout this paper) is to set all fields appearing in the $R$-equations to zero.
\begin{table}
\all{
\begin{array}{ c|lllllll }
\hline
\text{Equations} & \multicolumn{7}{c}{\text{Fields}}\\\hline
\multirow{2}{*}{$\{D_m, \G_0 \}$}
& & \nabla_m^{(4)} & \o_{mab}& H_{mna}& F_m& F_{mna} & F_{mabcd} \\
& &&&&& F_{abc} & \\\cline{2-8}
\multirow{2}{*}{$[D_a, \G_0]$ }
& && \o_{abm}&H_{mna}& F_m& F_{mna}& F_{mabcd}\\
& & & & H_{abc}& & F_{abc}& \\ \cline{2-8}
\multirow{2}{*}{$[D, \G_0 ]$}
&\p_m \phi  &  & & H_{mna}& F_m& F_{mna}&\\
 & &  & &H_{abc}& & F_{abc}& \\
 \hline \hline
\multirow{2}{*}{$\{D_a, \G_0 \}$}
&& \nabla_a^{(6)} & \o_{amn} & H_{mab} & F_a & F_{mab} & F_{a1234} \\
&& & & & & F_{mnp} & F_{mnabc} \\\cline{2-8}
\multirow{2}{*}{$[D_m, \G_0]$ }
& & & \o_{mna} & H_{mab} & F_a & F_{mab} & F_{a1234} \\
&&& & H_{mnp} & & F_{mnp}& F_{mnabc} \\ \cline{2-8}
\multirow{2}{*}{$\{D, \G_0\}$ }
& \p_a \phi & & & H_{mab} & F_a & F_{mab} & \\
& & &  & H_{mnp} & & F_{mnp} & \\ \hline
\end{array}
}
\caption{The string field components present in the various supersymmetry equations \eqref{n=4}.}
\label{tab1}
\end{table}

An important consequence of this simplifying ansatz is  that $\nabla_a^{(6)}$ acts trivially on the internal spinors. In the situation previously described, where we can explicitly identity a six-dimensional internal space $\mcal_6$, we thus require $\mcal_{6}$ to admit at least one covariantly constant spinor. Let us assume existence of some positive-chirality spinor $\xi^+$. Then the structure group of $\mcal_6$ reduces from $SO(1,5)$ to $SU(2) \ltimes \rbb^4$ \cite{gmr, figueroa, bryant}.
For a Riemannian manifold, a necessary condition for the existence of covariantly constant spinors is that the manifold is of special holonomy and Ricci flat, in particular, the internal space would have to be a Calabi-Yau threefold in the six-dimensional case. These conditions are no longer necessary for Lorentzian manifolds \cite{baum}. Nevertheless, simple examples which are of this type will suffice for our purposes, such as $\rbb^{1,5}$, $T^2 \times T^4$, $T^2 \times K3$ (potentially with a warped metric).
It would be possible to consider more sophisticated internal manifolds with non-constant spinors, which would be the analogue of generalizing the Calabi-Yau threefold to an $SU(3)$-structure space.
However, due to the fact that $L$-terms and $R$-terms are disjoint, any four-dimensional background that cannot be embedded by our methods is unlikely to be embeddable by considering more complicated internal spaces.\footnote{The sole reason why allowing $R$-terms might lead to more solutions is due to the fact that various components of the spin connection are not independent. This has not been relevant in our examples. }
The most enticing reason to examine more complicated backgrounds is that allowing more fluxes to be turned on leads to tunable parameters that one may wish to use in order to satisfy Bianchi identities.

\subsection{The $\ncal=1$ string supersymmetry equations}
In order to obtain the $\ncal = 1$ supersymmetry equations, we introduce a pair of Lorentzian D7-branes, whose supersymmetry requirements will place additional constraints on the ten-dimensional Killing spinor and thus reduce the number of supercharges preserved; we follow along the lines of  \cite{triendl, mpt}.
In flat local ten-dimensional coordinates $X^A$, we consider the case where the first D7-brane wraps the submanifold $\{ X^A \; | \; A \in \{0, ..., 7 \}\}$ and the second one wraps the submanifold $\{ X^A \; | \; A \in \{0, ..., 5, 8,9 \}\}$. We take both of them to have vanishing worldvolume flux; the open string supersymmetry conditions are thus given by
\eq{
\cg_{0567} \G_0 \ve = \cg_{0589} \G_0 \ve = \G_0 \ve = \ve \;,
}
leading to the identities
\eq{\label{g2chi2}
\cg_{05} \xi^\pm &=   \pm \xi^\pm\\
\cg_{67} \xi^\pm &= \pm i \xi^\pm\\
\cg_{89} \xi^\pm &= \pm i \xi^\pm \;.
}
Specifically, the D7-branes break the internal Lorentz symmetry
\eq{
SO(1,5) \rightarrow SO(1,1) \times SO(2) \times SO(2) \;.
}
From the four-dimensional point of view, the internal Lorentz symmetry corresponds to the R-symmetry. Hence by introducing the D7-branes we have reduced the R-symmetry group to  $\rbb_+ \times U(1) \times U(1)$, which does not act faithfully. We will come back to this later on. By making use of \eqref{g2chi2}, we can construct $2^3  = 8$ projection operators, such as for example $\frac12 \left(1 \pm \cg_{05}\right)$. So even after dropping the $R$-terms in \eqref{n=4}, when acted upon by every possible projection operator, the three $\ncal = 4$ supersymmetry conditions $L = L_m = L_a = 0$ give rise to  twenty-four equations. In order to reduce this to something more manageable, we will discard as many of these as we can. Specifically, the terms that are truly of interest are the four-dimensional Killing spinor equations. There are a number of fields appearing in these which correspond to auxiliary fields from the point of view of supergravity. We will set to zero any field that is algebraically unrelated to these fields. This leads to the following set of equations, which are equivalent to the total ten-dimensional IIB supersymmetry conditions (after taking into account that a number of components of the fluxes have been set to zero):\footnote{The derivation of these equation is very similar to what is explained in appendix C of \cite{mpt}. In comparison, here we have taken $\ve_+ = \ve$, $\ve_- = 0$, and we have shifted the dilatino equation by considering $D \rightarrow D - \G^a D_a$.}

\noindent The external gravitino leads to:
\eq{\label{extgrav}
 \left( \nabla_m + i A_m + i V_n \g^n \g_m  \right) \eta^+ - M^+ \g_m \eta^- &= 0\\
 \left( \nabla_m - i A_m - i V_n \g^n \g_m  \right) \eta^- - M^- \g_m \eta^+ &= 0\;.
}
The modified dilatino leads to:
\eq{\label{dilatino}
\left( \p_m \phi - \frac12 \O_m - 2 i V_m - 2 i B_m \right) \g^m \eta^+ + \left(- 4 M^+ + 2 i N^+\right) \eta^- &= 0 \\
\left( \p_m \phi - \frac12 \O_m + 2 i V_m + 2 i B_m \right) \g^m \eta^- + \left(- 4 M^- - 2 i N^-\right) \eta^+ &= 0 \;.
}
The internal gravitino leads to:
\eq{\label{intgrav}
\left(  i V_m  - \frac12 \O_{ma}^+ + 2i V_{ma} \right) \g^m \eta^+ + \left( M^+ + (  2i N_a^+ - 2 M_a^+)  \right) \eta^- &= 0 \\
\left(- i V_m  - \frac12 \O_{ma}^- - 2i V_{ma} \right) \g^m \eta^- + \left( M^- + (- 2i N_a^- - 2 M_a^-)  \right) \eta^+ &= 0 \;.
}
Let us first examine the external gravitino and the modified dilatino. We have defined the following fields in terms of the fluxes, dilaton and spin connection:
\eq{\label{n=1fields}
A_m       &=   \frac12          \left(i \o_{m05} + \o_{m67} + \o_{m89} \right) \\
\Omega_m  &=                    \o_{abm} \eta^{ab} \\
B_m       &=   \frac12          \left(i \o_{[05]m} + \o_{[67]m} + \o_{[89]m} \right) \\
V_m       &= - \frac18   e^\phi \left( F_m - F_{m6789} - i F_{m0567} - i F_{m0589} \right) \\
M^\pm     &=   \frac18 i e^\phi \left( F_{068} - F_{079} + i F_{569} + i F_{578} \mp \left(F_{568} - F_{579} + i F_{069} + i F_{078} \right) \right) \\
N^\pm     &=   \frac18 i        \left( H_{068} - H_{079} + i H_{569} + i H_{578} \mp \left( H_{568} - H_{579} + i H_{069} + i H_{078} \right) \right) \;.
}
The fields $(A, V, M^\pm)$ can be identified as the auxiliary fields of the $d=4$ $\ncal=1$ off-shell supergravity known as 16/16 supergravity \cite{16161, 16162, 16163}.\footnote{See appendix \ref{cd} for the explicit identification.} As the name suggests, 16/16 supergravity comes with 12+4 bosonic and fermionic degrees of freedom, whereas old and new minimal supergravity both come with 12. One can view 16/16 supergravity as a generalization of the two, either as old minimal supergravity coupled to a vector multiplet, or new minimal supergravity coupled to a chiral multiplet. In a way, this is the most natural candidate to expect. In order to construct field theory on curved spaces, one couples an appropriate supercurrent multiplet to the corresponding supergravity. From the respective degrees of freedom, it follows that the Ferrara-Zumino multiplet is coupled to old minimal supergravity, whereas the $\rcal$-multiplet is coupled to new minimal supergravity. Yet there are field theories which either do not admit an FZ-multiplet or an $\rcal$-multiplet, or admit neither. However, in all cases the non-minimal $\scal$-multiplet \cite{komargodski} does exist, which couples to 16/16 supergravity.

The ten-dimensional external gravitino supersymmetry conditions are precisely of the form of the Killing spinor equations imposed by the vanishing of the gravitino of 16/16 supergravity. However, obviously 16/16 supergravity was worked out in Lorentzian signature, whereas here, the metric is Riemannian. As a consequence the supercharges in our cases are $Spin(4)$ rather than $Spin(1,3)$, which requires that they are pseudoreal rather than real (or symplectic Majorana-Weyl rather than Majorana if one prefers). At the level of the supergravity action, this implies a doubling of the amount of supercharges. Indeed, this is reflected in the auxiliary fields: for 16/16 supergravity, the auxiliary fields $A$ and $V$ are required to be real, and $(M^-)^* = M^+$, whereas it is evident from \eqref{n=1fields} that no such requirement is imposed in our situation.

Similarly to the gravitino, the modified dilatino variation leads to the vanishing of the remaining fermion of 16/16 supergravity (which we will refer to as the dilatino from now on). In order to do so, one should eliminate the fields $(\O_m, B_m, N^\pm)$ which are not auxiliary fields of 16/16 supergravity. The scalars $N^\pm$ should be fixed as  $\pm i N^\pm = k M^\pm$ for some specific constant $k$. For appropriate metrics, the field $\Omega_m$ is exact and can be absorbed into $\d \phi$. The field $B_m$ is pure gauge and can be trivialized for any ten-dimensional metric. However, from the string theory perspective, there is no reason to impose any conditions whatsoever.

Finally, let us identify the fields appearing in the internal gravitino equations. The fact that some terms carry an internal index whereas others do not is a consequence of breaking internal Lorentz invariance by means of the D7-branes. The three-form fluxes appear through
\eq{\label{n=1fields2}
N^\pm_0 &=   \frac18 i \left(   H_{068} -   H_{079} \mp ( i H_{069} + i H_{078} ) \right)\\
N^\pm_5 &=   \frac18 i \left( i H_{569} + i H_{578} \mp (   H_{568} -   H_{579} ) \right)\\
N^\pm_6 &=   \frac18 i \left(   H_{068} + i H_{569} \mp (   H_{568} + i H_{069} ) \right)\\
N^\pm_7 &=   \frac18 i \left( - H_{079} + i H_{578} \mp ( - H_{579} + i H_{078} ) \right)\\
N^\pm_8 &=   \frac18 i \left(   H_{068} + i H_{578} \mp (   H_{568} + i H_{078} ) \right)\\
N^\pm_9 &=   \frac18 i \left( - H_{079} + i H_{569} \mp ( - H_{579} + i H_{069} ) \right)
}
and
\eq{
M_a^\pm[e^\phi F] = N_a^\pm[H]
}
satisfying
\eq{\label{nbreakup}
N_0^\pm + N_5^\pm &=  N_6^\pm + N_7^\pm = N_8^\pm + N_9^\pm = N^\pm \\
M_0^\pm + M_5^\pm &=  M_6^\pm + M_7^\pm = M_8^\pm + M_9^\pm = M^\pm \;.
}
In addition, we have defined
\eq{\label{n=1fields3}
\O_{m0}^\pm &=  - \o_{00m} \mp   \o_{05m} \\
\O_{m5}^\pm &=    \o_{55m} \pm   \o_{50m} \\
\O_{m6}^\pm &=    \o_{66m} \pm i \o_{67m} \\
\O_{m7}^\pm &=    \o_{77m} \mp i \o_{76m} \\
\O_{m8}^\pm &=    \o_{88m} \pm i \o_{89m} \\
\O_{m9}^\pm &=    \o_{99m} \mp i \o_{98m} \;.
}
and
\eq{\label{n=1fields4}
V_{m0} &= V_{m5} =\frac18 e^\phi   F_{m6789}  \\
V_{m6} &= V_{m7} =\frac18 e^\phi i F_{m0589}  \\
V_{m8} &= V_{m9} =\frac18 e^\phi i F_{m0567}
}
Unlike the external gravitino and modified dilatino equations, the internal gravitino equation has no four-dimensional interpretation. Instead, it should be considered as additional constraints imposed by demanding that a four-dimensional background dictated by the 16/16 supersymmetry conditions can be supersymmetrically embedded into a ten-dimensional background.

\section{Four-dimensional analysis}\label{sec:4d}
Having reduced the ten-dimensional supersymmetry conditions to  \eqref{extgrav}, \eqref{dilatino} and additional constraints originating from the internal gravitino \eqref{intgrav}, we now wish to see to what extent the known four-dimensional backgrounds can be lifted to solutions of the ten-dimensional supersymmetry constraints.

We will make use of known four-dimensional solutions to the Killing spinor equations \eqref{extgrav}, leading to specific backgrounds for supersymmetric field theories, and demonstrate how these are lifted to ten dimensions using our formalism. Such known solutions were constructed by coupling field theories to either old minimal \cite{st, df} or new minimal \cite{ktz, dfs} supergravity. The only fermionic field in both formulations is the gravitino, and hence coupling to minimal supergravity only requires imposing the vanishing of the gravitino variation.  The four-dimensional gravitino variation of minimal supergravities can be obtained from \eqref{extgrav} by imposing constraints on the auxiliary fields, specifically, $A=V$ for old minimal supergravity, $M^+ = M^- = 0$ for new minimal supergravity.
In contrast, coupling to 16/16 supergravity would require the (four-dimensional) dilatino variation to vanish as well. The combined supersymmetry fermionic variations of 16/16 supergravity are thus obtained by satisfying both  \eqref{extgrav} and \eqref{dilatino} and by imposing certain constraints on the auxiliary fields (see appendix \ref{cd}). As no four-dimensional analysis of field theories coupled to backgrounds of 16/16 supergravity is known, we will only examine minimal supergravity backgrounds. The lifting of these solutions to ten dimensions might either be direct or   pass via embedding into 16/16 supergravity.  All solutions in this paper satisfy the 16/16 conditions.

A number of prominent four-dimensional backgrounds are given on the topologies $T^{4-k} \times S^k$ and $T^{4-k} \times H_k$, where $H_k$ is $k$-dimensional hyperbolic space. Generically, the specific background auxiliary fields are not unique, and depending on which ones one chooses, a different number of supercharges is preserved by the theory. The specifics of the preserved supercharges define geometrical structures via the method of $G$-structures. Given a single nowhere vanishing chiral spinor, one can construct an $SU(2)$-structure consisting of an almost complex structure $J$ and a $(2,0)$- form $\O$. Given a pair of opposite chirality, one can construct a pair of such $SU(2)$-structures, the intersection of which is equivalent to a trivial structure, consisting of a pair of complex vector fields $(u, v)$. In the case of new minimal supergravity, the Killing spinor equations are such that the almost complex structures determined by the $SU(2)$-structures are integrable. One can introduce complex coordinates and a complex Killing vector $K$ can be found, by means of which the four-dimensional backgrounds can be characterized; see appendix \ref{cd} for the relation between $K$ and the trivial structure. In the case of old minimal supergravity, it turns out to be more convenient to work with the trivial structure instead. In our case, we will always make use of the trivial structure, even in some cases where it can only be introduced locally.\footnote{We understand that the notion of a $G$-structure describes global data of the underlying manifold, and that a `local' $G$-structure may thus appear rather ridiculous. However, this terminology is convenient to discuss local tensors which, if they were globally well-defined, would correspond to a $G$-structure.}

We would like to find ten-dimensional lifts of some known four-dimensional backgrounds.  The four-dimensional data
\eq{
(g_4, A, V, M^\pm, u, v, |\eta^-|/|\eta^+|)
}
are taken as input. These consist of the four-dimensional metric $g_4$ and the auxiliary fields $(A, V, M^\pm)$ associated to a solution of the Killing spinor equations \eqref{extgrav} for the metric $g_4$. As will be explained shortly, we find it convenient to specify the trivial-structure data $(u,v)$, as well as the relative spinor norm.

Given this four-dimensional data, we wish to define a supersymmetric string theory background
\eq{
(g_{10}, \phi, H, F_{1,3,5} )
}
such that the dilatino equations \eqref{dilatino} as well as the internal gravitino equations \eqref{intgrav} are solved and, by means of \eqref{n=1fields}, the auxiliary fields are indeed given by their specified values. Furthermore, we insist that the ten-dimensional Killing spinors $\ve^{1,2}$ are related to the four-dimensional supercharges $\eta^\pm$. This ensures a ten-dimensional lift of a minimal supergravity background.
As mentioned already, all of our solutions can also be viewed as lifts of 16/16 supergravity backgrounds.

Note that despite the fact that the gravitino variation of 16/16 supergravity allows for more solutions than minimal supergravity due to the presence of additional tunable auxiliary fields, the presence of additional constraints due to the dilatino variation means that, in general, the 16/16 Killing spinor equations appear to be more strict than those of than those of minimal supergravity.

We stress that the non-vanishing of the auxiliary fields means that the equations of motions of the corresponding supergravity are explicitly violated. This is irrelevant, as the goal of the procedure is to construct a new field theory Lagrangian. As a consequence, whether or not the supersymmetric string theory backgrounds we construct below solve the type IIB equations of motion is also beside the point. While our solutions to supersymmetry generally violate Bianchi identities, as we will see shortly, in some examples the Bianchi identities are explicitly solved. A supersymmetric background which satisfies the Bianchi identities solves the NSNS equations of motions for non-mixed spacetime indices; see \cite{pta} for a short review along the lines of \cite{lmmt}.\footnote{We work in the democratic type II formalism, where the Bianchi identities incorporate the RR-flux equations of motions.} Note that in all our solutions, we have set the $R$-terms in \eqref{n=4} to zero. By turning on the fields related to $\nabla_a^{(6)}$, which  contribute only to the $R$-terms, it might be possible to find deformations of our solutions for which the Bianchi identities are satisfied.

\subsection{Local trivial structure}\label{ssec:local}
Many four dimensional backgrounds are characterized by an $SU(2)$ structure and a (complex) Killing vector, which is tantamount to a trivial structure. The condition of its existence is equivalent to existence of a pair of nowhere vanishing spinors of opposite chirality. This condition is violated by some of the known four-dimensional backgrounds, notably $S^4$. Since all solutions presented here are local, and the trivial structure is a rather convenient tool, we shall make use of it to construct all our solutions. When the global conditions of existence of the trivial structure are not satisfied, the local solutions need to be properly extended.

Let us define a trivial structure following \cite{st}. Given two (local) normalized spinors $\hat{\eta}^\pm$ of $Spin(4)$ of opposite chirality, we can construct the following complex one-forms:
\eq{
v_m &=  \wt{\hat{\eta}^+} \g_m (\hat{\eta}^-)^c \\
u_m &=  \wt{\hat{\eta}^+} \g_m  \hat{\eta}^- \;.
}
Fierz identities lead to the conclusion that the four one-forms $v, v^*, u, u^*$ are orthogonal of norm two and trivialize the complexified cotangent bundle.
Furthermore, $\hat{\eta}^\pm$ satisfy
\eq{
\hat{\eta}^+ &= \frac12 v_m\g^m \hat{\eta}^- \\
\hat{\eta}^- &= \frac12 v_m^* \g^m \hat{\eta}^+
}
as well as the projection equations
\eq{\label{tsproj}
v_m   \g^m \hat{\eta}^+ &= u_m  \g^m \hat{\eta}^+ = 0\\
v_m^* \g^m \hat{\eta}^- &= u_m  \g^m \hat{\eta}^- = 0 \;.
}
Generically, the Killing spinors $\eta^\pm$ that solve \eqref{extgrav} need not be normalized. To compensate we have to deal with an additional factor
\eq{
\a = |\eta^-| / |\eta^+|
}
such that we have
\eq{\label{vdef}
\eta^+ &= \frac12 \a      v_m \g^m \eta^- \\
\eta^- &= \frac12 \a^{-1} v_m^* \g^m \eta^+ \;.
}
$\a$ is determined by the four-dimensional analysis. When $\scal$ admits a global trivial structure, as is the case with many examples, $\a$ can be eliminated altogether. When this is not the case, as for $S^4$, $\a$ and $\a^{-1}$ vanish at some points and the analysis is more involved.

\section{Non-compact backgrounds}
\label{sec:simple}
In this section we give a number of explicit string theory backgrounds which reproduce known four-dimensional backgrounds. As mentioned, this means solving the dilatino and internal gravitino equations \eqref{dilatino}, \eqref{intgrav} for a given solution of \eqref{extgrav}.  We  will expand the four-dimensional fields given in \eqref{n=1fields}, \eqref{n=1fields2}, \eqref{n=1fields3} and  \eqref{n=1fields4} in terms of the local trivial structure $(v, v^*)$ and  use \eqref{tsproj}, \eqref{vdef} to reduce the problem to a set of algebraic equations. Given such a solution, we can use the definitions of the four-dimensional fields to identify the ten-dimensional lifts. Note that in all cases, $H$ is such that the worldvolume flux $\fcal$ of the D3-brane can be set to zero, which is consistent with \eqref{kappa}.

All solutions presented in this section have two common features: they are non-compact, and preserve at least two supercharges of similar chirality. The latter feature is not just a technical assumption as we will explain in \ref{sec:complex}: backgrounds which for a given chirality preserve only one supercharge require a different approach.

\subsection{$H_4$}
We consider old minimal supergravity backgrounds on $H_4$. Such backgrounds allow for $(2,2)$ preserved supercharges \cite{st}. The four-dimensional background is given by
\eq{
g_4 &= \d \r^2 + e^{2 \r } ( \d x_1^2 + \d x_2^2 + \d x_3^2)  \\
M^\pm &= \frac12 \\
A &= V = 0 \;,
}
with the relevant data of the trivial structure being given by
\eq{
v &=  \d \r + i e^{ \r} \d x_3 \\
|\eta^-| / |\eta^+| &= 1\;.
}

Ten-dimensional fields can be written as
\eq{\label{h4background}
g_{10}     & = g_4 (x) + e^{2 \Delta (x)} g_6 (y) \\
\phi       &= 4  \Delta = 4  \r\\
H          &= 2 (e^{568} - e^{579}) \\
F_3        &= - 2 e^{- 4  \rho} \left( e^{569} + e^{578} \right)\\
F_1        &= F_5 = 0 \;.
}
We take the standard (warped) vielbeine for our metric, as explained in section \ref{connections}.
This reproduces the four-dimensional background auxiliary field profiles, and in addition sets
\eq{
\O_{ma}^\pm    &= \frac16 \Omega_m = \p_m \Delta \\
\pm i N^\pm_a  &= M^\pm_a \Rightarrow  \pm i N^\pm = M^\pm \\
V_{ma}         &=  0 \;.
}
Using these values, it follows that the dilatino equation \eqref{dilatino} and the internal gravitino equation \eqref{intgrav} are solved.
The Bianchi identities for $H$ and $F_3$ are not satisfied.

\subsection{$H_3 \times S^1$}
We consider new minimal supergravity backgrounds on $H_3 \times S^1$. The four-dimensional background is given by \cite{dfs}
\eq{\label{s3s1metric}
g_4   &= \d \t^2 + l^2 \left( \d \r^2 + e^{2 \r /l} ( \d x_1^2 + \d x_2^2) \right)  \\
A     &=  \frac{1}{2l} \d \t \\
V     &= -\frac{1}{2l} \d \t \\
M^\pm &= 0 \;,
}
and the trivial structure vector $v$ that can be deduced from the Killing vector $K$ is given by
\eq{
v &=  l \d \r + i \d \t \;.
}
Note that there is some freedom in $v$, in that we have fixed an arbitrary phase.
This background allows for $(2,2)$ preserved supercharges.

The ten-dimensional uplift is given by
\eq{
g_{10} &= g_4(x) + e^{2 \Delta(x)} g_6 (y) \\
\phi   &= 4 \Delta = 4 \r  \\
H      &= F_3 = F_5 = 0 \\
F_1    &=  \frac{4}{l} e^{-4 \r } \d \t \;.
}
This leads to the required four-dimensional profiles for the $g_4$ and the auxiliary field $V$. The auxiliary field $A$ is exact and is therefore generated by a gauge trasformation of the standard vielbeine, as described in appendix \ref{connections}.
In addition, we find
\eq{
\O_{ma}^\pm &= \frac16 \Omega_m = \p_m \Delta \\
N^\pm_a &= M_a = 0 \\
V_{ma} &= 0 \;.
}
Thus we conclude that \eqref{dilatino} and \eqref{intgrav} are solved. The Bianchi identities are not satisfied due to the fact that $\d F_1 \neq 0$. Reinstating the phase that we have fixed for the trivial structure alters the flux $F_1$, but it can be shown that no choice will lead to the Bianchi identities being satisfied.

\subsection{$S^3 \times \rbb$}\label{s3s120}
We consider a new minimal supergravity background on $S^3 \times \rbb$ with the standard metric and auxiliary fields given by \cite{dfs, shamir}
\eq{\label{s3s1input1}
g_4 &= \d \tau^2 + l^2 \left(\d \t^2 + \sin^2 \t \d \vt_1^2  + \cos^2 \t  \d \vt_2^2 \right) \\
A &=   \frac{i}{2l} \d \tau \\
V &= - \frac{i}{2l} \d \tau \\
M^\pm &= 0 \;,
}
and the relevant part of the trivial structure given by\footnote{Technically, we have that $K= \d \tau - i l \left(\cos^2 \t~ \d \vt_1 + \sin^2 \t ~\d \vt_2 \right)$. However, since all auxiliary fields are imaginary, the second external gravitino is invariant under complex conjugation up to $\xi^- \rightarrow (\xi^-)^c$. We thus have that $u \leftrightarrow v$, up to a (global) phase, with $u \sim K$. Furthermore, the external gravitino equations are invariant under change of global phase. Hence we can identify $K = v$. }
\eq{\label{s3s1input2}
v &= \d \tau - i l \left(\cos^2 \t~ \d \vt_1 + \sin^2 \t ~\d \vt_2 \right) \;.
}
This four-dimensional background leaves $(2,2)$ supercharges invariant. However, we will consider a string background that preserves $\ncal = (2,0)$ supersymmetry. This means that we decompose the ten-dimensional Killing spinor as
\eq{
\ve^1  = \xi^+ \eta^+ + \xi^{+c} \eta^{+c}  \;,
}
with $\eta, \eta^c$ the four-dimensional supercharges.
Although $S^3 \times S^1$ admits the existence of a negative chirality spinors as well, these do not appear in the decomposition of our Killing spinor $\ve^1$. We will however make use of these, by choosing an $\eta^-$ in such a way that we can construct the trivial structure as given in \eqref{s3s1input2}.

Ten-dimensional fields can be written as
\eq{
g_{10} &= g_4(x) + e^{2 \Delta_{05}(x)} g_2 (y) + e^{2 \Delta_{6789}(x)} \tilde{g}_4 (y)  \\
\phi &= 6 \Delta_{05} =  3 \Delta_{6789} = 6 \frac{\tau}{l} \\
H &= F_1 = F_3 = 0 \\
F_5 &=  \frac13 \d \exp\left(- 6 \tau \right) \wedge \left( e^{0567} + e^{0589} \right)  \;.
}
The internal space is chosen such that $\mcal_6 = \mcal_2 \times \mcal_4$, with the metric $g_2$ on $\mcal_2$ Lorentzian, and both $\mcal_2$, $\mcal_4$ allowing for covariantly constant spinors. The profiles of the fluxes reproduce the four-dimensional background auxiliary fields $(V, M)$. The auxiliary field $A$ is exact and can therefore be reproduced by taking a gauged $SO(1,1)$ rotation of the canonical vielbeine, as discussed in appendix \ref{connections}. In addition, the following need to satisfied:
\eq{
\O_{m0}^\pm &= \O_{m5}^\pm = \p_m \Delta_{05} \\
\O_{m6}^\pm &= \O_{m7}^\pm =  \O_{m8}^\pm = \O_{m9}^\pm = \p_m \Delta_{6789} \\
\Omega_m    &= \sum_{a} \O_{ma}^+ \\
N^\pm_a     &= M^\pm_a = 0 \\
V_{m6}      &= V_{m8} = \frac12 V_m \\
V_{m0}      &= 0 \;.
}
Using these, it follows that \eqref{dilatino} and \eqref{intgrav} are solved. If we take $\mcal_2$ and $\mcal_4$ flat we see that the Bianchi identities are explicitly satisfied. Lack of periodicity of the warp factors $\Delta_{05}$, $\Delta_{6789}$ as well as of the dilaton $\phi$ means that the solution cannot be compactified to $S^3 \times S^1$. It is possible to extend this background into a one-parameter family by splitting $\mcal_4$ into two two-dimensional components and taking warp factors $\Delta_{67} = k \Delta_{89}$ for some $k \in \rbb$, which can be compensated by means of the ratios between $F_{m0567}$ and $F_{m0589}$.

\section{Spherical backgrounds in complexified string theory}
\label{sec:complex}

There are a number of known four-dimensional backgrounds which cannot be obtained by the method outlined before. The most immediate reason is supersymmetry. The ten-dimensional Killing spinor $\ve$ is restricted to be Majorana-Weyl and thus is constrained to satisfy $\ve^c = \ve$. Since $Spin(4)$ spinors do not change chirality under charge conjugation, this means that $\ve$ can never be decomposed such that the four-dimensional part is determined solely by a single supercharge $\eta^+$, as the reality condition on $\ve$ immediately implies the necessity of also including a second supercharge of the same chirality $(\eta^+)^c$. Hence, the analysis done so far can apply to backgrounds with at least two preserved supercharges of the same chirality whereas backgrounds which admit only $(1,1)$ supercharges, such as $S^2 \times T^2$ and squashed $S^3 \times S^1$, are a priori not covered.

In order to  circumvent some of these problems we will use complexified  ten-dimensional supersymmetry, which  is obtained via  ``holomorphic complexification" as outlined in \cite{bergshoeff}.  Choosing a reality condition in this formulation then allows one to obtain the standard and the variant supergravities. The idea is to rewrite the type IIB action using only `holomorphic' fields, i.e., formally complexifying all the fields and not allowing complex conjugates  to appear anywhere. In particular, the Dirac conjugates in any spinor bilinears should be rewritten in terms of Majorana spinors. Indeed, one may note that the supersymmetry transformations only depend on such holomorphic quantities. As a result, one obtains a formulation of supergravity where  all reality conditions can be dropped;  bosonic fields are allowed to take values in $\mathbbm{C}$ and Majorana spinors are replaced by Dirac spinors. Crucially for our purposes, this means we allow for $\ve \neq \ve^c $.

A priori, the metric is allowed to take arbitrary complex values, while the self-duality condition of $F_5$ depends on the signature of the metric.
We will restrict our attention to backgrounds with real ten-dimensional  metric such that  there are no changes to the supersymmetry conditions \eqref{extgrav}, \eqref{dilatino}, \eqref{intgrav} due to the complexification of the fields.
In addition, reality of $\Gamma_0$ is required in order to derive \eqref{kappa}.\footnote{ For discussion of branes in variant supergravities see \cite{hull}.} We also bear in mind that there exists no consistent set of reality conditions on the fields that gives rise to  a real type IIB action with $(10,0)$ signature and therefore only consider ten-dimensional metrics which are Lorentzian.

Note that the complexifications of $\ncal=1$ supergravity and complex auxiliary fields have appeared in the four-dimensional analysis, so extending the range of ten-dimensional theories once again would mimic the four-dimensional analysis. At any rate, for backgrounds such as $S^2 \times T^2$ and squashed $S^3 \times S^1$ with $(1,1)$ supercharges, our method of obtaining a ten-dimensional lift works only for the complexified theory. On the other hand, once we are willing to consider such complexifications we may apply them to cases where this was not a priori needed, but for which we could not find ordinary ten-dimensional lifts. Both of the backgrounds we describe below,  $S^4$ and $S^3 \times S^1$, with $(2,2)$ supersymmetry, fall into this category.

\subsection{$S^4$}\label{secs4}
We consider old minimal supergravity backgrounds on $S^4$. The four-sphere allows for theories preserving more supersymmetry, but we will only consider (2,2) supercharges, consistent with our D7-brane setup. The metric and auxiliary fields of the background are given by
\eq{\label{s4}
g_4 &= \d \t^2 + \sin^2 \t \left( \d  \vt_3^2 + \sin^2 \vt_3 \left( \d \vt_2^2 + \sin^2 \vt_2 \d \vt_1^2 \right)\right)  \\
M^\pm &= \frac12 i \\
A &= V = 0 \;.
}
$S^4$ supports a nowhere vanishing spinor. It can be written as a sum of  chiral parts $\eta^+$ and $\eta^-$, which however necessarily have zeros (they do not vanish simultaneously). Also there are no nowhere vanishing vectors on $S^4$. The relevant {\sl local} trivial structure data can be written as \cite{st}
\eq{\label{s42}
v &= - i \d \t + \sin \t \left( \cos \vt_2 \d \vt_3 - \sin \vt_3 \sin \vt_2 \left( \cos \vt_3 \d \vt_2 + \sin \vt_3 \sin \vt_2 \d \vt_1 \right) \right) \\
\a &= \tan\left( \frac12 \t\right)\;.
}
Unlike all other backgrounds considered,  the norms of the Killing spinors $\eta^\pm$, measured by $\a = |\eta^-| / |\eta^+|$  are not identical. As mentioned, neither is non-vanishing, with zeroes at the north and south pole. This is another way of seeing  the four-sphere does not admit an almost complex structure, let alone an $SU(2)$-structure.

For the moment we shall restrict the range of $\t$ such that neither $\a$ not $\a^{-1}$ vanish, i.e. $0 < \t < \pi$ and find a local solution. Then we will extend the solution beyond this local patch. We present here the simplest solution; more elaborate versions can be found in appendix \ref{s4calc}.

Let us write the ten-dimensional fields as
\eq{\label{s4sol}
g_{10}   &= g_4 (x) + g_6 (y)  \\
\phi  &= 0 \\
H     &=  -2 i \left( e^{068} - e^{079} - i e^{569} - i e^{578} \right) \\
F_3   &= e^{068} - e^{079} - i e^{569} - i e^{578} \\
F_{1} &= - \frac{3}{2 \sin \t} v - \frac12 \cot\left( \frac12 \t\right) v^* \\
F_{5} &= \frac{1}{2 \sin \t} v \wedge \left( e^{6789} - i e^{0567} -i e^{0589} \right)+ \frac12 \cot\left( \frac12 \t\right) v^* \wedge e^{6789} \;.
}
As a result, the auxiliary fields are given by \eqref{s4}, and in addition
\eq{
\O_{ma}^\pm &= \O_m = 0 \\
N^\pm_a     &= \frac12 N^\pm = \pm 1 \\
M^\pm_a     &= \frac12 M^\pm = \frac14 i \\
V_{m0}     &= \frac{1}{2\sin \t} v_m  + \cot\left(\frac12 \t\right) v_m^*\\
V_{m6}      &= V_{m8}  = \frac{1}{2\sin \t} v_m   \;.
}
Plugging the above into \eqref{dilatino}, \eqref{intgrav}, we find that all string supersymmetry conditions are satisfied.
It can be checked that the worldvolume flux of the brane can be set to zero. The ten-dimensional fields do not satisfy the Bianchi identities.

With the exception of $F_5$ and $F_1$, the solution \eqref{s4sol} appears to be fine even beyond the range of validity $0 < \t < \pi$. By considering a slightly more complicated solution with a warp factor $e^{2 \Delta} \sim \sin^2 \t$, the problematic field can be tuned to be just $F_1 \sim \frac{1}{\sin \t} \d \t + ...$, with the ellipsis referring to regular terms (see appendix \ref{s4calc} for details on more general backgrounds on $S^4$). The simplest way of extending it is to first introduce the spherical stereographical coordinates $(R, \Theta_j)$ covering the sphere except for the north and south pole:
\eq{
\t &= 2 \arctan \frac{1}{R} \\
\vt_j &= \Theta_j \;.
}
Next, one rewrites these spherical stereographical coordinates in regular stereographical coordinates $Y_J \in \rbb^4 = S^4 \backslash\{(1,0,0,0,0)\}$, $J \in \{1,2,3,4\}$, thus covering the south pole as well and excluding only the north pole. Finally, to complete the covering, one considers the transition to another set of stereographical coordinates $\tilde{Y}_J \in S^4\backslash \{(0,1,0,0,0)\}$, which cover the sphere except for the ``east" pole. The two sets of stereographical coordinates are related as
\eq{
\ti{Y}_1 &= \frac{Y_J^2 - 1}{1 + Y_J^2 - 2 Y_1} \\
\ti{Y}_j &= \frac{2 Y_j}{1+ Y_J^2 - 2 Y_1} \;.
}
It is now not hard to verify that
\eq{
F_1 \sim \frac{\d \t } { \sin \t} = - \frac{\d R}{R} = - \frac{2 Y_J \d Y^J}{ Y_J^2}  = - \frac{2 \tilde{Y}_J \d \tilde{Y}^J}{ \tilde{Y}_J^2}
}
where all the coordinate sets are defined. Note that locally $F_1$ is constant, but it glues non-trivially across the patches, so is not closed, and hence does not satisfy the Bianchi identity.

\subsection{$S^3 \times S^1$ }
Let us revisit the $S^3 \times \rbb$ background as described in section \ref{s3s120} with metric and  auxiliary fields \eqref{s3s1input1} and trivial structure \eqref{s3s1input2}. Although we derived a ten-dimensional lift with (2,0) preserved supercharges, the modified dilatino equation \eqref{dilatino} and internal gravitino equation \eqref{intgrav} could not be satisfied if we allowed a spinor decomposition with (2,2) preserved supercharges; i.e., in the decomposition
\eq{
\ve^1 = \xi^+ \otimes \eta^+ + \xi^- \otimes \eta^- + \text{c.c.}
}
we were forced to take $\xi^- = 0$ (or $\xi^+ = 0$). In addition, the dilatino and warp factor were such that the background could not be compactified to $S^3 \times S^1$. By making use of complexified string theory, the fluxes allow for more degrees of freedom, which turn out to be sufficient to circumvent these issues.

In particular, we the following ten-dimensional uplift
\eq{
g_{10} &= g_4 (x) + e^{2 \Delta (x)} g_6(y) \\
\phi   &=  2 \Delta = \frac{1}{l} \varphi (\tau)  \\
H      &= - \frac{1}{2l}  \varphi' \left( e^{568} - e^{579} - i e^{069} - i e^{078} \right)\\&\phantom{=}
          - \frac{1}{l}            \left( e^{068} - e^{079} - i e^{569} - i e^{578} \right) \\
F_1    &=  \frac{16}{l} i\exp \left( - \frac{1}{l} \varphi \right) \d \tau \\
F_5    &=  \frac{2}{l}  \exp \left( - \frac{1}{l} \varphi \right) \d \tau \wedge \left( i e^{6789} + e^{0567} + e^{0589} \right) \\
F_3    &= 0 \;.
}
Here, $\varphi(\tau)$ is a scalar field to be chosen at will; in order to have a solution on $S^3 \times S^1$ rather than $S^3 \times \rbb$, one should take it periodic. Setting it to zero yields the most straightforward solution.
The fields have been constructed precisely such that $\forall a$
\eq{
\p_m \phi        &=  2 \O_{ma}^\pm = \frac{\varphi'}{2l} \left( v_m + v_m^* \right) = \frac{1}{l} \p_m \varphi  \\
\pm i N^\pm_a  &= \pm \frac12 i N^\pm = \frac{1}{4l} \left( \frac12 \varphi' \pm 1 \right)  \\
M_a^\pm        &= 0 \\
V_{ma}         &= - V_m = \frac{i}{4l} \left(v_m + v_m^* \right)
}
which solves both the dilatino equation \eqref{dilatino} and the internal gravitino equation \eqref{intgrav}. Regardless of choice of $\varphi$, the Bianchi identities are not satisfied, due to $d F_3 + H \wedge F_1 = H \wedge F_1 \neq 0$.

\subsection*{Acknowledgements}
We thank Hagen Triendl for collaboration in the early stages of this project. We would also like to thank Bruno Le Floch, Valentin Reys, Dimitrios Tsimpis and Alberto Zaffaroni for useful discussions. R.M  would like to thank KIAS for hospitality during the conclusion of this work. This work was supported in part by the Agence Nationale de la Recherche under the grant 12-BS05-003-01 (R.M.) and by ERC Grant Agreement n. 307286 (XD-STRING) (D.P.).

\clearpage
\appendix

\section{Conventions and Killing spinor equations}\label{cd}
Since this paper relies heavily on using known four-dimensional results, let us discuss how to convert our four-dimensional conventions to those in the literature. In this section, and this section only, use will be made of two-component spinor indices $\a, \b, ...$ and $\dot{\a}, \dot{\b}, ...$. Since everything in this appendix is four-dimensional, these should not be confused with the curved indices of the six-dimensional internal space.

In general, all our conventions are identical to those used in \cite{mpt}. Let us also mention the following identities which are needed here but not yet mentioned in \cite{mpt}: in order to obtain the $\ncal = 1$ supersymmetry conditions, the following gamma-matrix (anti-)commutator identities are needed, which can be obtained by making use of the D7-brane identities \eqref{g2chi2}:
\eq{\label{Xcom2}
-\frac14 \{\{X^{ab} \cg_{ab}, \cg_{67} \}, \cg_{89} \} \xi^\pm &=  \pm 2!i  \left(i X_{05}  + X_{67}  + X_{89} \right) \xi^\pm  \\
-\frac14 [[X_{a} \cg^{a}, \cg_{67}], \cg_{89}] \xi^\pm&=  0\\
-\frac14 [[X_{abc} \cg^{abc}, \cg_{67}], \cg_{89}] \xi^\pm &=
                                        - 3!\Big(  X_{068} - X_{079} + i X_{569} + i X_{578} \\
&\phantom{=} \mp \left(X_{568} - X_{579} + i X_{069} + i X_{078}\right)             \Big) \cg_{068} \xi^\pm
}
and
\eq{\label{Xcom3}
-\frac14 [[X^{ab} \cg_{ab}, \cg_{67}], \cg_{89}]       \xi^\pm &= - 2 \left( X_{68} - X_{79} \mp i  ( X_{69} + X_{78})\right) \cg_0 \cg_{068} \xi^\pm\\
-\frac14 \{\{X_{a} \cg^{a}, \cg_{67}\}, \cg_{89}\}     \xi^\pm &= - (X_0 \pm X_5 ) \cg_0 \xi^\pm \\
-\frac14 \{\{X_{abc} \cg^{abc}, \cg_{67}\}, \cg_{89}\} \xi^\pm &=  -3! i \left( X_{567} + X_{589} \pm (X_{067} + X_{089}) \right) \cg_0 \xi^\pm \;.
}

\subsection{Minimal supergravity}
The Killing spinor equations of new minimal supergravity are given by \cite{dfs}
\eq{\label{nm}
\left( \nabla_m - i \ti{A}_m + i \ti{V}_m + i \ti{V}^n \s_{mn} \right) \eta_\a &= 0 \\
\left( \nabla_m + i \ti{A}_m - i \ti{V}_m - i \ti{V}^n \s_{mn} \right) \eta^{\dot {\a}} &= 0 \;,
}
with $\eta = (\eta_\a , \eta^{\dot{\a}})$. In terms of such two-component spinors, our charge conjugation matrix and gamma matrices are defined as
\eq{\label{gammasigma}
\g_m = \left( \begin{array}{cc}
0 & i (\s_m)_{\a\dot{\a}}\\
i (\tilde{\s}_m)^{\dot{\a}\a} & 0
\end{array}\right) \;,
\qquad
C =
- \left( \begin{array}{cc}
i (\s_2)_\a^{\phantom{\a}\b} & 0 \\
0 & i (\s_2)^{\dot{\a}}_{\phantom{\a}\dot{\b}}
\end{array} \right)\;,
}
with $(\s_m)_{\a\dot{a}} = ( \vec{\s}, -i)$ and $(\tilde{\s}_m)^{\dot{\a}\a} = (- \vec{\s}, -i)$. The resulting gamma matrices are Hermitian and satisfy
$\g_{1234} = - \text{diag}(1, -1)$, hence $\eta_\a$ is of negative chirality and $\eta^{\dot{\a}}$ is of positive chirality.
We therefore see that the Killing spinor equations \eqref{extgrav} reduce to the new minimal ones \eqref{nm} if we set
\eq{
V_m &= - \frac12 \tilde{V}_m \\
A_m &= \tilde{A}_m - \frac12 \tilde{V}_m \\
M^\pm &= 0 \;.
}
Furthermore, use is often made in the literature on new minimal supergravity backgrounds of the vector-field $K^\sharp$ and associated one-form $K$. Generically, new minimal backgrounds require the metric to be Hermitian, and hence one can choose local complex coordinates $(w, z)$ in terms of which
\eq{
g_4 &= \L^2 \left[ \left( \d w + h(z) \d z\right) \left( \d \bar{w} + \bar{h} \d \bar{z} \right) + c(z, \bar{z})^2 \d z \d {\bar z} \right] \\
- i u &=  \L \left( \d \bar{w} + \bar{h} \d \bar{z} \right) \\
- i v &=  \L c  \d \bar{z} \;.
}
In particular, using $K^\sharp = \p_w$, one notes that $u \sim K$ up to a real normalization factor.\footnote{Provided that $M^\pm  = 0$, one can always change the global phases of $\eta^\pm$ to transform the global phases of $(u,v)$ without changing the auxiliary fields.}

The Killing spinor equations of old minimal supergravity are given by \cite{st}
\eq{\label{om}
\left( \nabla_m + 2 b_m + b^n \g_{nm} \right) \eta^+  + M \eta^- &= 0 \\
\left( \nabla_m - 2 b_m - b^n \g_{nm} \right) \eta^-  - \ti{M} \eta^+ &= 0 \;.
}
We therefore see that the Killing spinor equations \eqref{extgrav} reduce to the old minimal ones \eqref{om} if we set
\eq{
A_m &= V_m = -i b_m \\
M^+ &= - M \\
M^- &= \ti{M} \;.
}

\subsection{16/16 supergravity}
The Lorentzian two-component formulation of the vanishing of the gravitino and the additional fermion of 16/16 supergravity is given by \cite{16161}
\eq{\label{1616fermionicvariations}
\delta \psi_{m \a} &= \nabla_m \z_\a - \frac12 i n \bar{S} \s_{m \a\dot{\a}} \bar{\z}^{\dot{\a}}
                     - i \left( \frac{1}{2n} G_m - \frac{3n + 1}{4} e^{4n \psi}  W_m \right) \z_\a
                     + \frac12 i \left( \s_m \cdot \bar{\s}_n \right)_\a^{\phantom{\a}\b} G^n \z_\b \\
\delta T_\a &= - \frac14 S \z_\a + \frac12 \left( \frac{1}{2n} G_m - \frac{1}{4n} e^{4n \psi}  W_m - i \p_m \psi \right) \s^m_{\a\dot{\a}} \bar{\z}^{\dot{\a}} \;.
}
Here, $n \in \rbb \backslash \{[-\frac13, 1]\}$ is a free parameter, $T_\a$ is the dynamical fermion of the theory, $G_m, W_m$ are one-forms, and we should stress that $\psi$ is a scalar. We have taken the fermionic terms in the variations to vanish. Roughly speaking, we wish to identify a `Euclidean version' of \eqref{1616fermionicvariations} with the supersymmetry conditions \eqref{extgrav}, \eqref{dilatino}. By this, we mean that we will allow complex values for the auxiliary fields $G_m, W_m$, and will not identify $\bar{S}$ with the complex conjugate of $S$.

Using \eqref{gammasigma}, we see that the gravitino variation matches \eqref{extgrav} if we identify
\eq{\label{identification1}
V_m &= - \frac12 G_m \\
A_m &= \left(\frac{1}{2n} + 1\right) G_m - \frac{3n+1}{4} e^{4n \psi} W_m \\
M^- &= \frac12 n \bar{S} \;.
}
The variation of the ten-dimensional modified dilatino $\chi$ as given in \eqref{dilatino} contains more fields than the variation of the four-dimensional $T_\a$. Therefore, some of the ten-dimensional fields have to be fixed in a certain way to ensure that the ten-dimensional modified dilatino can be identified with the four-dimensional fermion. Ten-dimensional backgrounds for which the fields are not fixed this way are not lifts of 16/16 backgrounds. Specifically, using \eqref{identification1}, $\chi = k T_\a$ for some constant $k$ implies
\eq{
- 4 M^+ + 2 i N^+ &= -\frac{k}{4} S \\
\p_m \phi - \frac12 \O_m - 2 i B_m &= - \frac{k}{2} \left( \p_m \psi + i \left[ \left(\frac{1}{2n} + \frac{2}{k} \right) G_m - \frac{1}{4n} e^{4n \psi} W_m \right]\right) \;.
}
The former equation can always be satisfied. Since the fields $(G_m, W_m)$ can be identified with $(A_m, V_m)$, it is not generically guaranteed that the latter equation is satisfied for an arbitrary profile of the ten-dimensional fields. However, in all examples of backgrounds given in this paper, $(A_m, V_m, \O_m)$ are exact (and $B_m=0$), which translates to exactness of $(G_m, W_m)$. In such cases, the latter equation can be considered a defining equation for $\p_m \phi$ in terms of $\p_m \psi$. Therefore, all our backgrounds correspond to lifts of 16/16 backgrounds.

\section{Metrics, spin connections and R-symmetry}\label{connections}
In this section we will outline ten-dimensional metrics that are used, the corresponding spin connections, and the resulting profiles for the fields $A$, $B$, $\O$ and $\O_{ma}^\pm$ which appear in the supersymmetry conditions \eqref{extgrav}, \eqref{dilatino}, \eqref{intgrav} and are defined in \eqref{n=1fields}, \eqref{n=1fields2}. In essence, we will always make use of (warped) block diagonal metric. Due to the breaking of the internal $SO(1,5)$ Lorentz symmetry by the D7-branes, we occasionally break down the internal metric to three two-dimensional pieces with different warpings. The reason is that such block diagonal metrics are the simplest possibility. However, it is not too much more difficult to extend this to metrics with off-diagonal components. For the sake of completeness, we will discuss the most general case as well.

The most general ten-dimensional metric is given by
\eq{\label{mostgenmetric}
g_{10} = g_{\m\n}(x,y) \d x^\m \d x^\n + 2 \acal_{\m\a}(x,y) \d x^\m \d y^\a + g_{\a\b}(x,y) \d y^\a \d y^\b \;.
}
Associated to this metric, we can define the vielbein
\eq{\label{vbgauge}
e_M^A = \left( \begin{array}{cc}
e_\m^m(x,y) & \acal_\m^a (x,y) \\
0 & e_a^\a (x,y)
\end{array} \right) \;, \quad
e^M_A = \left( \begin{array}{cc}
e^\m_m(x,y) & - \acal^\a_m (x,y) \\
0 & e^\a_a (x,y)
\end{array} \right)\;,
}
leading to spin connections
\eq{\label{fiberedsc}
\o_{\m ab} &=  e^\a_{[a} \p_\m e_{b]\a}  - \acal^{\a}_\m e^{\b}_{[a} \p_\a e_{b]\b} - e^\a_{[a} e_{b]\b} \p_\a \acal_{\m}^\b \\
\o_{ab \m} &=  e^\a_{(a} \p_\m e_{b)\a}  - \acal^{\a}_\m e^{\b}_{(a} \p_\a e_{b)\b} - e^\a_{(a} e_{b)\b} \p_\a \acal_{\m}^\b \\
\o_{amn}   &=  e^\m_{[m} \p_a e_{n]\m}  - e^\m_{[m} e^\n_{n]} e_{\a a}\left(\p_\m - \acal_\m^\b \p_\b \right) \acal_\n^\a    \\
\o_{mna}   &=  e^\m_{(m} \p_a e_{n)\m}  + e^\m_{[m} e^\n_{n]} e_{\a a}\left(\p_\m - \acal_\m^\b \p_\b \right) \acal_\n^\a \;.
}
Clearly, the above is not invariant under exchange of internal and external indices, whereas the metric \eqref{mostgenmetric} is. This is due to the choice of the vielbein \eqref{vbgauge}.

Let us now restrict attention to the case of interest. Setting $\acal =0$,  the components of the spin connection reduce to
\eq{\label{bdsc}
\o_{mab} &=  e^\a_{[a} \p_m e_{b]\a} \\
\o_{abm} &=  e^\a_{(a} \p_m e_{b)\a} \\
\o_{amn} &=  e^\m_{[m} \p_a e_{n]\m} \\
\o_{mna} &=  e^\m_{(m} \p_a e_{n)\m} \;.
}
Generally, we will be interested in the subcases
\eq{
g_{10} = g_4(x) + e^{2 \Delta(x)_{05} }  g_2 (y) + e^{2 \Delta(x)_{67} }  \tilde{g}_2 (y) + e^{2 \Delta(x)_{89} }  \hat{g}_2(y)
}
and
\eq{
g_{10} = g_4(x) + e^{2 \Delta(x) }  g_6 (y) \;.
}
The latter allows more freedom for the internal space, but the former, where we split the six-dimensional metric into a block diagonal one comprised of a Lorentzian metric (in the flat 0, 5 directions) and two Riemannian metrics (respectively in the flat 6,7 and the flat 8,9 directions), allows more freedom for the auxiliary fields, due to the various warp factors which depend on the external coordinates. To be precise, we should introduce different indices for all three internal two-dimensional metrics, but to avoid cluttering the notation, we will not do so.

Given the above metrics, one may introduce diagonal vielbeine $e_\m^m (x)$, $e^{\Delta_{(a)}} e_\a^a(y)$ which square to the metric and simplify the spin connection \eqref{bdsc} to
\eq{\label{bdsc2}
\o_{amn} &= \o_{mna} = 0 \\
\o_{mab} &= 0 \\
\o_{abm} &= \eta_{ab} \p_m \Delta_{(a)} \;,
}
where $\Delta_{(a)} = \Delta_{05} (\delta_{a0} + \delta_{a5}) +   \Delta_{67} (\delta_{a6} + \delta_{a7}) + \Delta_{89}  (\delta_{a8} + \delta_{a9})$
with no summation over $a$.  With respect to the supersymmetry conditions, the first line is always necessary for our ansatze to hold (see table \ref{tab1}), the second line dictates the auxiliary field $A$, the third line determines $B$, $\O$ and $\O_{ma}^\pm$ (see \eqref{n=1fields}, \eqref{n=1fields2}).

However, the choice of vielbein is not unique; any local $SO(1,1) \times SO(2) \times SO(2)$ rotation of the internal vielbeine $e_\a^a(y)$ (acting on the flat indices of the vielbein) leaves the metric invariant, but not the spin connection.
From the four-dimensional point of view, such transformations are gauged R-symmetry transformations. As an example, let $\t$ be a coordinate on a 1-cycle on the Riemannian external space $\scal$. Then we see that if we consider the vielbeine
\eq{\label{rotations}
\hat{e}_\a^0 &= \left( \cosh k_2 \t e_\a^0 - \sinh k_2 \t e_\a^5 \right) \\
\hat{e}_\a^5 &= \left(-\sinh k_2 \t e_\a^5 + \cosh k_2 \t e_\a^5 \right) \\
\hat{e}_\a^6 &= \left(  \cos k_1 \t e_\a^6 - \sin  k_1 \t e_\a^7 \right) \\
\hat{e}_\a^7 &= \left(  \sin k_1 \t e_\a^6 + \cos  k_1 \t e_\a^7 \right) \;,
}
with arbitrary $k_{1,2} \in \rbb$ the metric remains invariant (as well as $\o_{abm}, \o_{amn}$ and $\o_{mna}$), yet the change in spin connection $\o_{mab}$ leads via \eqref{n=1fields} to
\eq{
A \rightarrow A + \frac12 (k_1 + i k_2) \d \t \;.
}
This is, as expected, a gauge transformation for the R-symmetry gauge field $A$. In this way, exact background profile for the auxiliary field $A$ can be constructed. We shall do so frequently.

Given the spin connection \eqref{bdsc}, it follows from \eqref{n=1fields} that $B=0$. In fact, this holds even for the spin connection \eqref{fiberedsc}
which is valid for the most general metric. Therefore, the field $B$ is pure gauge. However, it might still be necessary for certain solutions to use a non-zero $B$. This can be achieved by making use of a different choice of vielbeine for a generic ten-dimensional metric: since the metric \eqref{mostgenmetric} does not distinguish between external and indices, we can consider vielbeine where we have swapped the indices:
\eq{\label{vbgauge2}
e_M^A = \left( \begin{array}{cc}
e_\a^a(x,y) & \acal_\a^m (x,y) \\
0 & e_\m^m (x,y)
\end{array} \right) \;, \quad
e^M_A = \left( \begin{array}{cc}
e^\a_a(x,y) & - \acal^\a_m (x,y) \\
0 & e^\m_m (x,y)
\end{array} \right)\;.
}
The spin connection components are given by
\eq{
\o_{\a mn} &=  e^\m_{[m} \p_\a e_{n]\m} - \acal^{\m}_\a e^{\n}_{[m} \p_\m e_{n]\n} - e^\m_{[m} e_{n]\n} \p_\m \acal_{\a}^\n \\
\o_{mn \a} &=  e^\m_{(m} \p_\a e_{n)\m} - \acal^{\m}_\a e^{\n}_{(m} \p_\m e_{n)\n} - e^\m_{(m} e_{n)\n} \p_\m \acal_{\a}^\n \\
\o_{mab}   &=  e^\a_{[a} \p_m e_{b]\a}  - e^\a_{[a} e^\b_{b]} e_{\m m}\left(\p_\a - \acal_\a^\n \p_\n \right) \acal_\b^\m   \\
\o_{abm}   &=  e^\a_{(a} \p_m e_{b)\a}  + e^\a_{[a} e^\b_{b]} e_{\m m}\left(\p_\a - \acal_\a^\n \p_\n \right) \acal_\b^\m \;.
}
As a result, $\o_{[ab]m}$ is no longer automatically vanishing, hence such a choice of vielbein gauge can be used to construct a non-zero auxiliary field $B$ appearing in \eqref{dilatino}. None of our examples make use of this possibility.

\section{More general $S^4$ backgrounds}\label{s4calc}
In this section, the procedure to find string backgrounds embedding a four-sphere wrapped by a D3-brane is given. In order to solve \eqref{dilatino}, \eqref{intgrav}, making use of the $S^4$ data \eqref{s4}, \eqref{s42}, we will take the following ans\"{a}tze for the $\ncal = 1$ fields:
\eq{\label{s4ansatz}
\O_{0}^\pm &= \frac12 i f_0 \left(v - v^* \right) \mp \frac12 i g_0 \left(v - v^* \right)\\
\O_{5}^\pm &= \frac12 i f_5 \left(v - v^* \right) \pm \frac12 i g_0 \left(v - v^* \right) \\
\O_{\hat{a}}^\pm &= \frac12 i f_{\hat{a}} \left(v - v^* \right)  \\
V_{a} &= \frac{1}{8} \left( h_a v + \ti{h}_a v^* \right) =
\frac{1}{8} \left( [ \tilde{h}_a +\frac12 (h_a - \ti{h}_a)]  \left(v+ v^*\right)  + +\frac12 (h_a - \ti{h}_a)  \left(v- v^*\right) \right)\\
N^\pm_0 &= \frac12 N^\pm \pm \a^{\pm 1} S^\pm \\
N^\pm_5 &= \frac12 N^\pm \mp \a^{\pm 1} S^\pm \\
M_a^\pm &= \frac12 M^\pm \\
\d \phi &= \frac12 i \varphi \left(v - v^*\right) \\
\tilde{\varphi} &= \varphi - \frac12 \sum_a f_a\\
B &= 0 \;.
}
Here, $\hat{a}, \hat{b}, ... \in \{6,7,8,9\}$. Note that it is always possible to absorb a shift of $N_a$ into $M_a$, a freedom we will not make use of.
Plugging these into \eqref{dilatino} and \eqref{intgrav}, making use of \eqref{tsproj}, \eqref{vdef} and \eqref{nbreakup}, one ends up with the following algebraic equations:
\eq{
2 (N^+ + N^-) &=
- \left( \a + \a^{-1} \right) f_0 + \left( \a - \a^{-1} \right) g_0 - \a \left(\ti{h}_0 + 2 S^+ \right) - \left(h_0 + 2 S^- \right)\\
&= - \left( \a + \a^{-1} \right) f_{\hat{a}}  - \a \ti{h}_{\hat{a}} - \a^{-1} h_{\hat{a}}\\
&=  \left( \a + \a^{-1} \right) \ti{\varphi} \\
\phantom{}\\
2 (N^+ - N^-) &=
   - \left( \a - \a^{-1} \right) f_0 + \left( \a + \a^{-1} \right) g_0 - \a \left(\ti{h}_0 + 2 S^+ \right) + \left(h_0 + 2 S^- \right)\\
&= - \left( \a - \a^{-1} \right) f_{\hat{a}}  - \a \ti{h}_{\hat{a}} + \a^{-1} h_{\hat{a}}\\
&=   \left( \a - \a^{-1} \right) \ti{\varphi} + 4 \\
\phantom{}\\
S^- &= S^+ - g_0 \\
f_0 - f_5 &= 2 (g_0 - 2 S^+) \;.
}
We solve a number of these equations by setting $f_6 = f_7 = f_8 = f_9$, $h_{6} = h_{8}$, $\ti{h}_6 = \ti{h}_8$.
Using a few trig identities to solve the rest leads to the following supersymmetry constraints on string backgrounds containing a kappa-symmetric four-sphere:
\eq{
\frac12 \left(h_{\hat{a}} - \ti{h}_{\hat{a}} \right) &= \frac{2}{\sin \t} \\
\frac12 \left(h_0 - \ti{h}_0 \right) &= \frac{2}{\sin \t}  \\
\varphi &= 2 f_{\hat{a}} - 2 \cot \left( \frac12 \t \right)- \ti{h}_0 \\
N^\pm &= \pm 1 + \frac12 \a^{\pm 1} \left( - f_5 - g_0 + 2 S^+  - 2 \cot \left(\frac12 \t \right) - \ti{h}_0 \right) \\
S^- &= S^+ - g_0 \\
f_0 - f_5 &= 2 (g_0 - 2 S^+) \;.
}
The parameters $(\tilde{h}_0, f_5, f_6, g_0, S^+)$ are all free, and given any such set, it is then possible to compute the string fields leading to such a background, and to construct a ten-dimensional metric reproducing the right spin-connection components (notably, $g_0$). Since both $\frac12 \left(h_0 - \ti{h}_0 \right)$ and $\frac12 \left(h_{\hat{a}} - \ti{h}_{\hat{a}} \right)$ are singular outside of $\theta \in (0, \pi)$, there is an issue when trying to extend the solution to a globally well-defined solution. From the field definitions \eqref{n=1fields}, \eqref{n=1fields4} and \eqref{s4ansatz}, it follows that $F_{m0567}$ scales as $\frac{1}{\sin \t}$; it is possible to compensate this by taking the warp factor  to scale as $\Delta \sim \log \left(\sin \t \right) + ...$.
However, this trick cannot be played with $F_m$, which is constrained in terms of $F_5$ by $V_m = 0$. It thus follows that the solution is not globally well-defined.

The relatively simple solution given in section \ref{secs4} corresponds to the choice
\eq{
g_0      &= S^+        =  f_a = 0 \\
\ti{h}_0 &=  2 \cot \left(\frac12 \t\right) \;.
}

\end{document}